\documentclass[epj]{svjour}
\usepackage{latexsym}
\usepackage{graphics}
\usepackage{graphicx}% Include figure files
\usepackage{amsmath}
\usepackage{lipsum}
\usepackage{multirow}
\usepackage{footnote}
\usepackage{units}
\usepackage{epstopdf}
\usepackage{epsfig}
\usepackage{lpic}
\usepackage{color}
\usepackage[makeroom]{cancel}
\begin{document}\sloppy

\title{Design and performance of an absolute $^3$He/Cs magnetometer}
%\subtitle{Do you have a subtitle?\\ If so, write it here}
\author{H.-C. Koch\inst{1}\inst{2} \and G. Bison\inst{3} \and Z. D. Gruji\'c\inst{1} \and W. Heil\inst{2} \and M. Kasprzak\inst{1} \and P. Knowles\inst{1}\thanks{\emph{Present address:} LogrusData, Vienna, Austria} \and A. Kraft\inst{2} \and A. Pazgalev\inst{5} \and A.\,Schnabel\inst{4} \and J. Voigt\inst{4} \and A. Weis\inst{1}% etc
% \thanks is optional - remove next line if not needed
%\thanks{\emph{Present address:} LogrusData, Vienna, Austria}%
}                     % Do not remove
%
%\offprints{}          % Insert a name or remove this line
%
\institute{Physics Department, University of Fribourg, CH-1700, Fribourg, Switzerland \and Department of Physics, Johannes Gutenberg-University, D-55122 Mainz, Germany \and Paul Scherrer Institute, CH-5232, Villigen, Switzerland \and Physikalisch-Technische Bundesanstalt, Berlin, Germany \and Ioffe Physical Technical Institute, Russian Academy of Sciences, 194021 St. Petersburg, Russia}
\date{Received: \today / Revised version: date}
% The correct dates will be entered by Springer
%
\abstract{
 % end of PACS codes
\noindent We report on the design and performance of a highly sensitive
combined $^3$He/Cs magnetometer for the absolute measurement of
magnetic fields.
The magnetometer relies on the magnetometric detection of the free
spin precession of nuclear spin polarized $^3$He gas by
optically pumped cesium magnetometers.
We plan to deploy this type of combined magnetometer in an
experiment searching for a permanent electric dipole moment of ultracold
neutrons at the Paul Scherrer Institute (Switzerland).
A prototype magnetometer was built at the University of Fribourg (Switzerland) and tested at
Physikalisch-Technische Bundesanstalt (Berlin, Germany).
We demonstrate that the combined magnetometer allows Cram\'er-Rao-
limited field determinations with recording times in the range of
$\unit[10\sim 500]{s}$, measurements above $\unit[500]{s}$ being limited by the stability of
the applied magnetic field.
With a $\unit[100]{s}$ recording time we were able to perform an absolute measurement of a magnetic field of $\approx\unit[1]{\mu T}$ with a
standard uncertainty of $\Delta B\sim\unit[60]{fT}$, corresponding to $\Delta B/B<$6$\times$10$^{-8}$.
} %end of abstract
\maketitle
%
%%%%%%%%%%%%%%%%%%%%%%%
\section{Introduction}
\label{sec:n2edm}
%%%%%%%%%%%%%%%%%%%%%%%
%
\noindent A new experiment searching for a permanent  electric dipole
moment of the neutron (nEDM) is currently being developed at Paul Scherrer
Institute (PSI), Switzerland \cite{Baker:2011:SNE}.
In the experiment, the degeneracy of the neutron's magnetic sublevels is lifted by the interaction of the neutron's magnetic moment $\vec\mu_n=g_I \mu _N\vec{I}/\hbar\equiv \mu _n\vec{I}/I$ with a static magnetic field $\vec{B}_0$, where $\vec{I}$ is the neutron's angular momentum.
Additionally, a static electric field $\vec{\mathcal{E}}$ is applied parallel or antiparallel to $\vec{B}_0$.
In case the neutron has an EDM  $\vec d_n=d_n\vec{I}/I$,
the external field interaction Hamiltonian is given by
\begin{equation}
H_{ext}=-\frac{\mu_n}{I} \vec I \cdot \vec{B}_0 -\frac{d_n}{I}\vec{I} \cdot \vec{\mathcal E}\,.
\end{equation}
The Larmor precession frequencies 
\begin{align}
\omega_{\uparrow \uparrow}=\left|\frac{\mu_n B_{0\uparrow\uparrow} - d_n \mathcal E}{I\hbar}\right|&\,&\text{and}&\,&\omega_{\uparrow \downarrow}=\left|\frac{\mu_n B_{0\uparrow\downarrow} + d_n \mathcal E}{I\hbar}\right|
\end{align}
in transverse (perpendicular to $\vec I$) magnetic fields are given by the sublevel splitting
for parallel ($\uparrow\uparrow$) and antiparallel ($\uparrow\downarrow$) magnetic and electric fields \cite{Pendlebury:2004:GPI}.
In the PSI experiment the precession frequency of spin-polarized ultracold neutrons
is measured by Ramsey's method of (time-) separated oscillatory fields \cite{ramsey1950molecular}.
A measurement cycle takes $\approx\unit[400]{s}$ and consists of filling the
neutron storage vessel, running a Ramsey cycle, emptying the vessel and measuring the
neutrons' spin polarization.
The nEDM experiment measures whether or not the (magnetic) spin precession frequency is altered by an electric field applied along the magnetic field.

In practice one compares the precession frequencies in experiments with ($\uparrow\uparrow$) and ($\uparrow\downarrow$) configuration.
From the difference frequency 
\begin{equation}
\label{eq:edm}
\omega_{\uparrow\uparrow}-\omega_{\uparrow\downarrow}=\frac{2d_n}{\hbar}\left(\mathcal{E}_{\uparrow\uparrow}+\mathcal{E}_{\uparrow\downarrow}\right)+\frac{2\mu_n}{\hbar}\left(B_{0\uparrow\uparrow}-B_{0\uparrow\downarrow}\right)\,,
\end{equation}
the nEDM is inferred.
From Eq.~\eqref{eq:edm} it is clear that the magnetic field has to be precisely controlled and known during the precession time of the neutrons (typically $\sim\unit[180]{s}$), hence the need for very accurate magnetometers.
An uncorrected statistically fluctuating difference $B_{\uparrow\downarrow}-B_{\uparrow\uparrow}\neq0$ would worsen the statistics for the nEDM experiment since the electric-field-induced difference in precession frequencies (if any) is many orders of magnitude smaller than the Larmor frequency itself.
Even more severe, a change in the magnitude of the applied magnetic field that is correlated to the electric field direction would be misinterpreted as an nEDM, if not corrected for.  
Moreover, magnetic field gradients may lead to geometrical phase effects \cite{Pendlebury:2004:GPI} that could also mimick an nEDM.
A high-sensitivity and high-accuracy measurement of
the magnetic field and its variation over the neutron storage volume during the free evolution time is therefore of crucial
importance for controlling and suppressing several major systematic errors and ensuring good statistics in the nEDM experiment.
%
%

%%%%%%%%%%%%%%%%%%%%%%%%%%%%%%%%%%%%%%%%%%%%%%%%%%%%%%%%%%
\section{Magnetometry in the PSI-EDM experiment}
%%%%%%%%%%%%%%%%%%%%%%%%%%%%%%%%%%%%%%%%%%%%%%%%%%%%%%%%%%
%
\noindent The current nEDM experiment at PSI\cite{Baker:2011:SNE} deploys two types of optically pumped magnetometers for measuring the temporal and spatial variations of the magnetic field in and around the neutron storage volume, viz., a Hg comagnetometer and an array of 16 Cs magnetometers (CsOPMs).
We first note that the Hg and the Cs magnetometers as well as the magnetometer described in this work are all scalar magnetometers that measure only the modulus of the magnetic field vector at the sensor's location.
The working sensitivities of the currently deployed systems (Cs, Hg) are both $\sim\unit[100]{fT}$ in a $\unit[100]{s}$ measurement time.
This is sufficient for the ongoing phase of the project since the statistical sensitivity of the nEDM measurement is currently limited by the neutron counting statistics.
The uncertainties arising from this limitation currently allow a tolerance of magnetic field fluctuations up to $\approx\unit[100]{fT}$ during one Ramsey cycle without loss of sensitivity.
The Hg co-magnetometer  \cite{fertl2013laser} yields a volume-averaged value of the field in the neutron bottle, important for normalization of the neutron precession frequency.
The co-magnetometer uses (nuclear) spin-polarized $^{199}$Hg vapor that occupies the same storage volume as the neutrons, and the magnetic field is inferred from the frequency of the free spin precession (FSP) of the Hg's spin polarization.
The Hg-FSP is monitored by recording the (time dependent) transmitted power of a resonant circularly polarized  light beam traversing the Hg vapor.
Strictly speaking, the spin precession of the Hg atoms is only `quasi'-free, since the read-out light beam may affect the spin precession frequency (e.g., by the light shift effect \cite{kastler1963displacement}), thereby limiting the magnetometer's accuracy.
The relative systematic shift arising from this effect for the given parameters of the current nEDM experiment has been estimated to be on the order of $6\times 10^{-9}$ \cite{fertl2013laser}.
A limiting factor to the accuracy of field measurements with the Hg magnetometer is the Hg gyromagnetic ratio, which is only known with a relative accuracy of $\sim 10^{-6}$ \cite{cagnac1960orientation},\cite{mohr2012codata}.

The Cs magnetometer array builds on an optically detected magnetic resonance process \cite{Groeger:2006:HSL}, in which the frequency of a weak applied oscillating magnetic field is made identical (`locked') to the Cs atoms' Larmor precession frequency using a feedback loop \cite{Knowles:2009:LDC}.
The Cs magnetometers offer the possibility to access information on the spatial field distribution outside the UCN precession volume, needed to control magnetic field gradients.
The Cs magnetometers are also prone to possible light shift effects, but their accuracy may be more seriously affected by phase errors (and the stability thereof) in their feedback electronics.
Moreover, because of the hyperfine interaction, the Cs magnetometer readings are affected by the quadratic Zeeman effect and, being driven magnetometers (in contrast to magnetometers based on free precession), their interaction with the magnetic resonance driving rf field introduces a systematic frequency shift (Bloch-Siegert shift \cite{Bloch:1940:MRN}).
The systematic effects mentioned above lead to sensor-specific offsets of the magnetometric readings that are generally unknown, not necessarily constant and may depend on other experimental parameters.
This spoils the direct comparison of the absolute field value given by different sensors and limits the ability to obtain a consistent picture of the magnetic field inside the apparatus necessary to suppress systematic errors in the EDM measurement such as those mentioned in Sec.~\ref{sec:n2edm} caused by the geometric phase effects~\cite{Pendlebury:2004:GPI}.

\noindent The combined $^3$He/Cs magnetometer described hereafter offers an important complement to the magnetometers discussed above.
It is based on recording the FSP of nuclear spin polarized $^3$He gas, by detecting the time dependent magnetic field produced by the precessing (and decaying) $^3$He magnetization with an arrangement of several Cs magnetometers.
$^3$He FSP detection by external magnetometers provides an indirect optical readout nonperturbative to the $^3$He FSP, avoiding possible systematic effects as they may occur by the read-out beams or the feedback control in the Hg and Cs magnetometers, respectively.
In the absence of magnetic field gradients the Larmor precession frequency of $^3$He is thus an absolute measure of the magnetic field inside the magnetometer cell.
When gradients are present, the measured precession frequency corresponds to the volume averaged field in the cell.
The details of this averaging process will depend on the dynamic regime in the cell \cite{Cates:RelInhom}.
On the other hand, the systematic effects affecting the CsOPMs are irrelevant for their use as readouts for $^3$He FSP.
The fact that the $^3$He gyromagnetic ratio is known with a relative precision of 2.5$\times$10$^{-8}$ \cite{mohr2012codata} makes $^3$He a promising candidate as a reference for magnetic field measurements, as suggested in \cite{Flowers:1993:MNM}.
One application that is envisioned for the next stage of the nEDM experiment (n2EDM) is a $^3$He-``quasi comagnetometer''.
Since in this experiment it is not possible to have $^3$He cohabiting inside the neutron bottle, flat cylindrical magnetometer vessels will be installed above and below the cylindrical neutron precession chamber.
These vessels will have the same geometrical cross section as the precession chamber and will thus be traversed to first order by the same magnetic flux.
An array of CsOPMs around the vessels will be used to detect the $^3$He FSP signal.
The magnetic field measurement from the $^3$He/Cs magnetometer, performed simultaneously with the nEDM measurement proper, can then be used, along with the Hg co-magnetometer, to normalize the neutron precession frequency and to correct for field changes \cite{Harris:1999:NEL},\cite{Kraft2014}.
From the double-chamber type of geometry, additional information on magnetic field gradients can be obtained.
Another possible application is an array of several compact $^3$He/Cs magnetometers to measure the magnetic field at different spatial positions around the precession chamber.
Since $^3$He/Cs magnetometers do not suffer from the systematics discussed above, and offsets which they may induce, readings from different $^3$He/Cs sensors can be more easily compared and their use can thus improve field stabilization and gradient control in the experiment. 
In this paper, we describe a prototype $^3$He/Cs magnetometer based on a small, sealed spherical $^3$He cell surrounded by Cs magnetometers.
The prototype is not adapted to later use in the nEDM experiment, but designed as a versatile test device for investigation of the combined magnetometer concept. 

%%%%%%%%%%%%%%%%%%%%%%%%%%%%%%%%%%%%%%%%%%%%%%%%%%%%%%%%%%%%%%
\subsection{Preparation and detection of polarized $^3$He}
\label{sec:prepareDetect}
%%%%%%%%%%%%%%%%%%%%%%%%%%%%%%%%%%%%%%%%%%%%%%%%%%%%%%%%%%%%%%%%
%
\noindent $^3$He gas in a spherical glass cell is polarized by optical pumping of the helium atoms in the  metastable 2$^3$S$_1$ state populated by collisions in a weak glow discharge, using circularly polarized 1083~nm light from a 2W  ytterbium-doped fiber laser.
The electronic spin polarization is transferred to nuclear spin polarization of the groundstate atoms by metastable exchange.
Metastable exchange optical pumping (MEOP) is a method that is well studied and described in detail, e.g.,  in \cite{Happer:1972:OP} and \cite{leduc2000kinetics}.
Conversely to Cs and Hg magnetometers, the spin polarization (and hence its dynamics) cannot be detected by optical means because of the large optical excitation energy (20~eV) of the $^3$He ground state for which no (convenient) light sources are available.

The rotating magnetization $\vec{m}_\mathrm{He}$ that is associated with the precession of the nuclear spin polarization can be detected by pick-up coils \cite{colegrove1963polarization}, which are very inefficient at the low precession frequency ($\sim\unit[30]{Hz}$) occuring here.

SQUIDs (superconducting quantum interference devices) are highly sensitive magnetometers that have been deployed for detecting the $^3$He FSP \cite{Gemmel2010}.
However, the additional technical complexity associated with the cryogenic cooling needed for the SQUID operation is an obstacle for operating these magnetometers under the experimental conditions of a room temperature nEDM experiment.
In 1967, Cohen-Tannoudji and co-workers \cite{Cohen-Tannoudji:1969:DSM} have demonstrated the suitability of a discharge lamp pumped alkali (Rb) vapor magnetometer for detecting the FSP of $^3$He nuclei.
Based on the results reported in that paper, we estimate their magnetometric sensitivity to be $\approx \unit[80]{pT \cdot s^{3/2}/T_m^{3/2}}$, assuming a Cram\'er-Rao limited performance (c.f.~Sec.~\ref{sec:anal}) in a measurement time $T_m$.

%%%%%%%%%%%%%%%%%%%%%%%%%%%%%%%%%%%%%%%%%%%%%%%%%%%%%%%%%%%%%%
\subsection{$^3$He/Cs magnetometer principle}
\label{sec:principle}
%%%%%%%%%%%%%%%%%%%%%%%%%%%%%%%%%%%%%%%%%%%%%%%%%%%%%%%%%%%%%%%%
%
%
\noindent During optical pumping the $^3$He cell is exposed to a homogeneous magnetic field $\vec{B}_0$ oriented along the pump laser beam.
The oriented nuclear magnetic moments give rise to a macroscopic magnetization $\vec{m}_\mathrm{He}$ that produces a magnetic dipole-like field $\vec{B}_\mathrm{He}$ outside of the cell.
One readily estimates that the field from a $\unit[100]{\%}$ polarized gas at $\unit[1]{mbar}$ is on the order of $\unit[200]{pT}$ on the outside surface of the cell.
After optical pumping, the laser and the gas discharge are turned off and a $\pi/2$ rf-pulse is applied to the cell in order to flip $\vec{m}_\mathrm{He}$ to a plane perpendicular to $\vec{B}_0$, upon which $\vec{m}_\mathrm{He}$ starts freely precessing around  $\vec{B}_0$ at the $^3$He Larmor frequency
\begin{equation}
\label{eq:principle_larmor}
\omega_\mathrm{He}=\gamma_\mathrm{He}\left|\vec{B}_0\right|\,,
\end{equation}
where
$\gamma_\mathrm{He}/2\pi=\unit[32.43410084(81)]{Hz/\mu T}$ \\($\Delta\gamma_\mathrm{He}/\gamma_\mathrm{He}=$2.5$\times$10$^{-8}$) is the gyromagnetic ratio of the $^3$He nucleus \cite{mohr2012codata}.
The precessing (and decaying) magnetization produces at the position $\vec{r}_\mathrm{Cs}$ of each Cs magnetometer a magnetic field $\vec{B}_\mathrm{He}(\vec{r}_\mathrm{Cs},t)$ with time-dependent amplitude and orientation. 

\begin{figure}
\includegraphics[width=0.97\columnwidth]{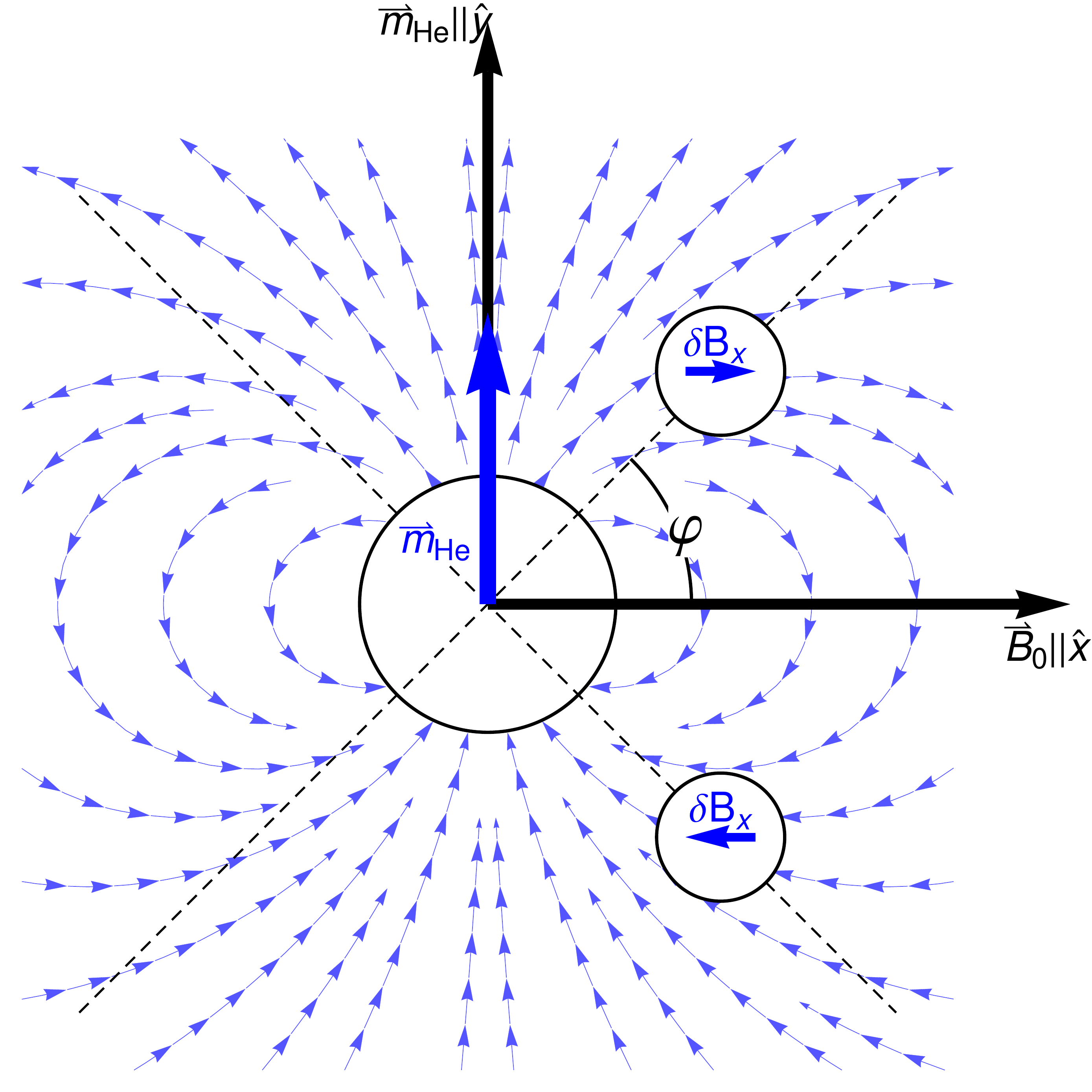}
\caption{$^3$He spin sample cell with its magnetic dipole field in the ($\hat{\vec{x}}$,$\hat{\vec{y}}$)-plane. The direction of the applied $\vec{B}_0$ defines the $\hat{\vec{x}}$ axis, so that the $^3$He magnetization precesses in the ($\hat{\vec{y}}$,$\hat{\vec{z}}$) plane. The Cs sensors shown at their different positions record the FSP signal phase-shifted by $\pi$. The $45^\circ$ cones of highest sensitivity are denoted by the dashed lines.}
\label{fig:principle_gradio} 
\end{figure}

The Cs magnetometers are scalar magnetometers, i.e., they measure the modulus
$B(\vec r,t)=\left|\vec{B}_0(\vec r,t)+\vec{B}_\mathrm{He}(\vec r,t)\right|$
of the total field at their location.
Since $B_\mathrm{He}{\ll}B_0$, and $B_0$ is nominally constant in time one has $B\approx B_0+\hat{\vec{B}}_0\cdot \vec{B}_\mathrm{He}(t)$,
so that the CsOPMs are, to first order, only sensitive to the component $\delta B_x$ of the $^3$He-FSP field along the applied magnetic field $\vec{B}_0$.
A simple calculation shows that---for a given distance $r_\mathrm{Cs}$---this time dependent projection has a maximum amplitude when the sensors are located on a double cone with a half-opening angle of $\varphi$=45$^\circ$ with respect to $\vec{B}_0$.
In the prototype described below, the centers of all 8 CsOPM cells were located on that double cone, and the relative azimuthal positions on the cones determine the phase relations between the individual FSP signals detected  by the different CsOPMs (compare Fig.~\ref{fig:principle_gradio}).
By pairwise subtraction of CsOPM signals that are dephased by $\pi$, common mode magnetic noise components (such as magnetic fields oscillating at the 50~Hz line frequency) that are in-phase on both sensors can be strongly suppressed in the differential signal, while increasing the signal of interest, as shown in Sec.~\ref{sec:common}.
For the chosen geometry of the combined magnetometer, 8 such gradiometer pairs can be formed.

%%%%%%%%%%%%%%%%%%%%%%%%%%%%%%%%%%%%%%%%%%%%%%%%%%
\subsection{The Cs magnetometers}
%%%%%%%%%%%%%%%%%%%%%%%%%%%%%%%%%%%%%%%%%%%%%%
%

\begin{figure}
\centering
\includegraphics[width=0.97\columnwidth]{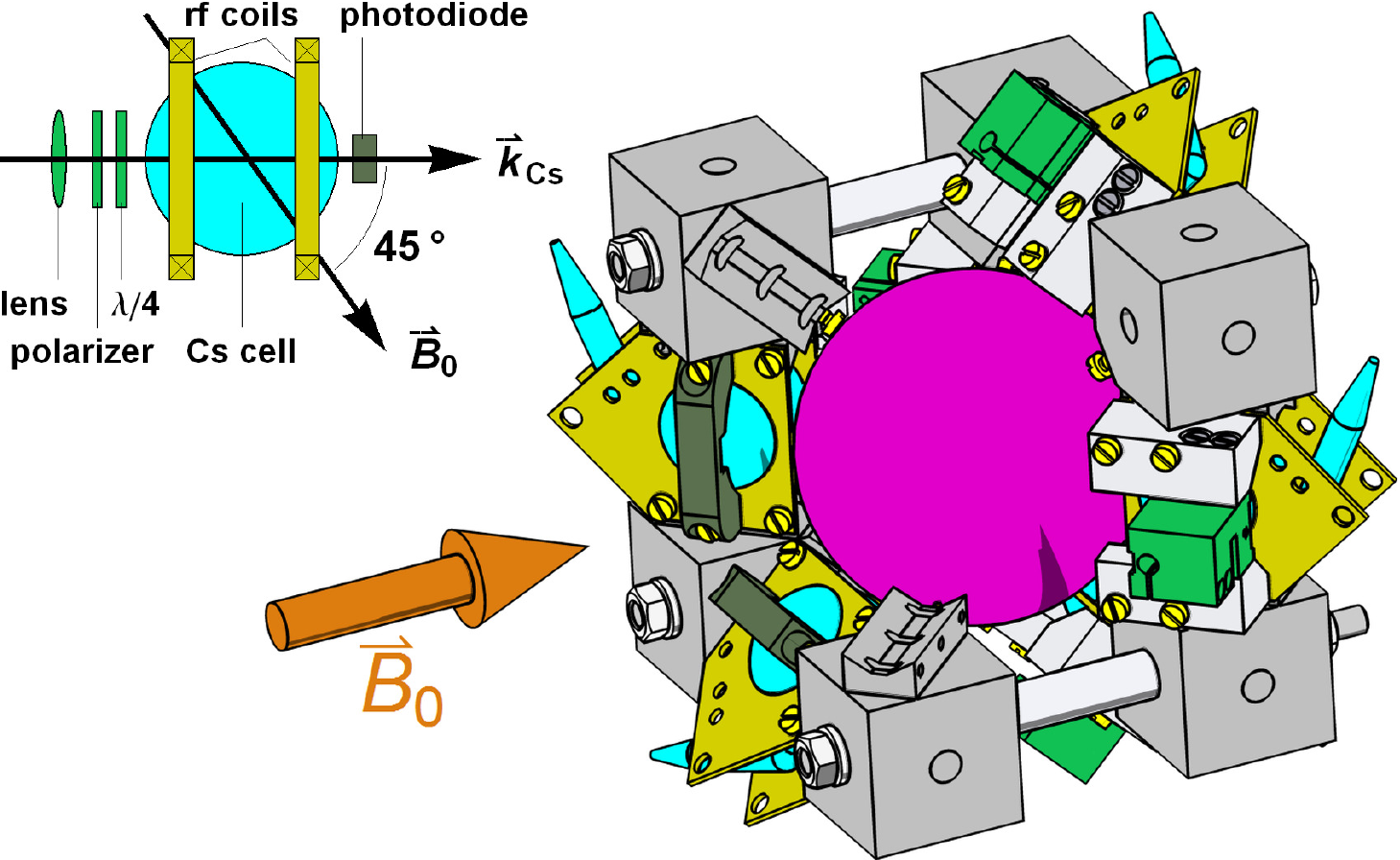}
\caption[Short Caption]{(Color online) Schematic drawing of the CsOPM (left) and CAD drawing of the combined $^3$He/Cs magnetometer (right). The spherical $^3$He cell (magenta) in the middle of the cubic structure is surrounded by eight CsOPMs (blue) on the edges of the cube, in which the rf coils are laid out on printed circuit boards (yellow). The total dimensions of the combined magnetometer are $\sim (\unit[10]{cm})^3$. One corner cube and two CsOPMs are left out for better visibility.}
\label{fig:magnetometer}
\end{figure}

\begin{figure}
\hspace*{-22mm}
  \includegraphics[width=1.5\columnwidth]{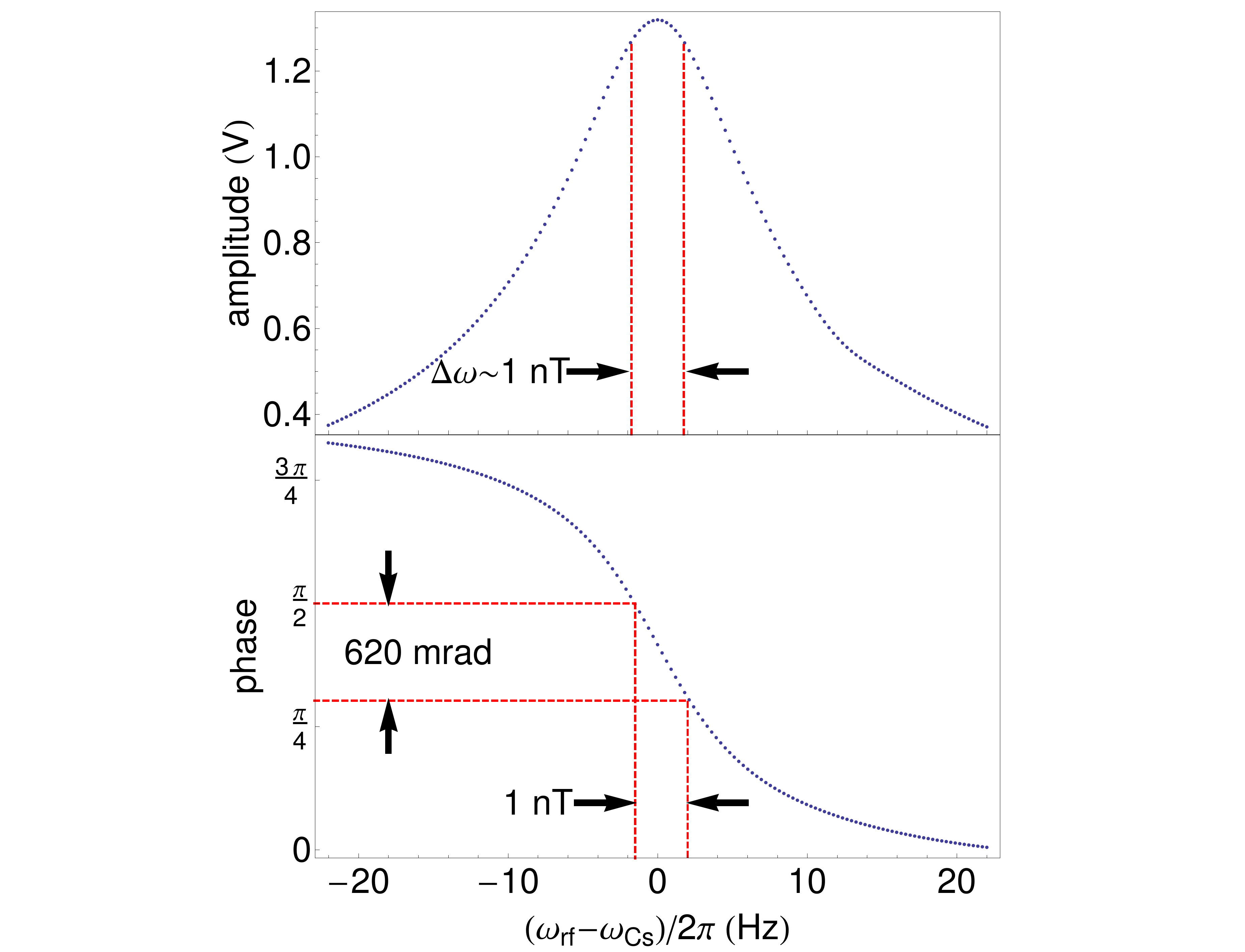}
\caption[Short Caption]{Amplitude (top) and phase (bottom) response of a CsOPM when sweeping the rf frequency $\omega_\mathrm{rf}$. The dashed lines represent the change of the CsOPMs Larmor frequency $\omega_{Cs}$ and phase response corresponding to a 1~nT variation of the magnetic field $B_0$.}
\label{fig:response}
\end{figure}

\noindent The CsOPMs used in this study are laser pumped double-resonance magnetometers, operated in the $M_x$ configuration \cite{Groeger:2006:HSL}.
A sketch of a CsOPM is shown in Fig.~\ref{fig:magnetometer} (left).
Light from a Toptica DLPro diode laser with a frequency actively stabilized to the $F_g{=}4{\rightarrow}F_e{=}3$ transition of the Cs-D$_1$ line at 894~nm is delivered to each magnetometer module via a multimode fiber.
The light from each fiber is collimated and given a circular polarization, after which it passes through a room-temperature paraffin-coated \cite{castagna:Cells:2009} 30~mm diameter Cs vapour cell.
The cell coating ensures a long-lived coherence ($\sim\unit[30]{ms}$) of the spin polarization created by optical pumping.
The light exiting the cell is detected by a photodiode (PD) that measures the Cs vapor's optical transmission.
The propagation direction $\hat{\vec{k}}_{Cs}$ of the incident light makes an angle of 45$^\circ$ with respect to $\vec{B}_0$ for all CsOPMs since this yields maximal sensitivity \cite{Groeger:2006:HSL}.
A weak magnetic field (rf field) parallel to $\vec{k}_{Cs}$ oscillating at frequency $\omega_\mathrm{rf}$ resonantly drives the precession of the Cs vapour's magnetization, thereby modulating the vapour's absorption coefficient and hence the photodiode signal \cite{Groeger:2006:HSL}.
The (transimpedance-) amplified PD signal is demodulated by a dual channel digital lock-in amplifier referenced to $\omega_\mathrm{rf}$.
Figure \ref{fig:response} shows the dependence of the amplitude and phase of a CsOPM sensor on the detuning from the resonance $\delta=\omega_\mathrm{rf}-\omega_{Cs}$.
The resonance occurs at the Cs Larmor frequency $\omega_\mathrm{rf}=\omega_\mathrm{Cs}=\gamma_\mathrm{Cs}\left|\vec{B}_0\right|$, where $\gamma_\mathrm{Cs}\approx 2\pi\times\unit[3.5]{kHz/\mu T}$ is the Cs ground state's gyromagnetic ratio.
In the currently employed mode of operation in the nEDM experiment the rf frequency is tuned near the line center where the phase has a linear dependence on the frequency detuning and the phase signal is used to drive a voltage-controlled oscillator generating the oscillatory voltage for the rf coils.
This mode of operation represents a feedback loop that keeps $\omega_\mathrm{rf}$ locked to the Cs Larmor frequency \cite{Groeger:2006:HSL}.
When multiple CsOPMs are operated in close spatial vicinity one sensor may be parasitically driven by the rf of a neighboring magnetometer.
To avoid this effect known as cross talk, it is advisable to drive all CsOPMS at the same common frequency.
In the measurements described here the CsOPMs were thus driven at a single, constant frequency $\omega_\text{rf}$ close to $\omega_{Cs}$ and the photodiode signals were demodulated at that fixed rf-frequency using digital lock-in amplifiers.
The oscillatory magnetic field $B_\mathrm{He}(t)$ then leads to an oscillation of the CsOPM's phase signal at $\omega_\mathrm{He}$, with an amplitude proportional to $B_\mathrm{He}$, as long as $B_\mathrm{He}\gamma_\mathrm{Cs}{\ll}\Delta\omega$, where $\Delta\omega$ is the linewidth of the resonance of Fig.~\ref{fig:response}.
We note that the proportionality factor between the phase response of the CsOPM and $B_\mathrm{He}$ depends on the Cs cell properties and the parameters of operation.
For the CsOPMs driven at fixed frequency, bandwidth limitations arising from the lifetime of the Cs polarization and the lock-in demodulator filter have to be considered.
The scaling of the phase signal to magnetic units which we will do in the following to make the results more intuitively accessible does not correct for these effects, e.g., the reported amplitudes do not reflect the true values of $B_\mathrm{He}$.
The true FSP amplitudes are actually roughly a factor of five larger.
%%%%%%%%%%%%%%%%%%%%%%%%%%%%%%%%%%%%%%%
\section{The prototype magnetometer}
\label{sec:magnetometer}
%%%%%%%%%%%%%%%%%%%%%%%%%%%%%%%%%%%%
%
\noindent A prototype of a combined $^3$He/Cs magnetometer for studying the $^3$He FSP readout with laser pumped CsOPMs was built in Fribourg.
It consists of a $\unit[70]{mm}$ diameter spherical cell filled with $^3$He at a pressure of $\unit[1]{mbar}$ that is fixed in the center of a mechanical structure holding eight laser-pumped CsOPMs mounted symmetrically on the edges of a cube as shown in Fig.~\ref{fig:magnetometer} (right), thereby fulfilling the optimal sensitivity criterion (45$^\circ$ cone) discussed above.
The distance between the $^3$He cell center and the Cs cell centers is $\unit[50.5(5)]{mm}$.
The He cell carries two electrodes, each consisting of a spiral of copper foil glued to the outside of the cell that are driven by a $\unit[1.2]{MHz}$ sinusoidal voltage (amplified by a Tesla transformer) to ignite and sustain a weak gas discharge in the cell.
The electrodes were designed to achieve a homogeneous illumination of the cell volume by the gas discharge while allowing optical access to a large part of the cell's surface to permit the $^3$He pump laser beam to traverse the cell.

Figure~\ref{fig:magnetometer} (left) shows details of a single CsOPM sensor.
Each sensor carries its own pair of rf coils and the coils of all 8 sensors were driven at the same, constant rf frequency during measurements.
This mode of operation is only possible when the magnetic field gradients over the whole structure are sufficiently small so that the individual Larmor frequencies of all 8 sensors differ by amounts that are much less than the Cs magnetic resonance linewidth $\Delta\omega$.
During the measurements described here, the local Larmor frequencies of two CsOPMs $i,j\in \left\{ 1,8\right\}$ differed by less than $(\omega_i-\omega_j)/(2\pi)<\unit[1.8]{Hz}$.

The apparatus depicted in Fig.~\ref{fig:magnetometer} (right) is surrounded by large coils (Helmholtz configuration, $\sim\unit[30]{cm}$ diameter, not shown in the figure) for producing magnetic fields perpendicular to $\hat{\vec{B}}_0$.
They were used to start the FSP by flipping the $^3$He magnetization by resonant rf-pulses following the pumping process.
Having in mind the later mounting of the device in the vacuum chamber of the nEDM spectrometer at PSI, all components were manufactured from nonmagnetic and vacuum compatible (low outgassing) materials.

%%%%%%%%%%%%%%%%%%%%%%%%
\section{Measurements}
\label{sec:measure}
%%%%%%%%%%%%%%%%%%%%%%%%
%
\noindent A key issue of the present study was the determination of the intrinsic magnetometric sensitivity of the combined $^3$He/Cs magnetometer.
Since the stability of the applied magnetic field sets a limit on the ability to determine the magnetometer sensitivity (c.f.~Sec.~\ref{sec:compare} for details), that field has to be kept as stable as possible.
After initial tests in Fribourg the measurements reported below were carried out in the magnetically shielded room BMSR-2 \cite{Thiel:2007:PTB-Demagnetization} at the Physikalisch-Technische Bundesanstalt (PTB) in Berlin, Germany.
The BMSR-2 room  is one of the magnetically most quiet and stable places on earth.
It consists of a 7~layer MU-metal magnetic shield and an additional aluminum layer. 
It features a built-in multi-channel system of (vector) SQUID magnetometers that were operated together with our test equipment.
A three axis Helmholtz coil system \cite{Hilschenz:Coilsystem2010} was used to produce the $B_0$ field inside BMSR-2.
The coil was driven by a commercial low-noise current source (Magnicon, CSE-1) delivering a current of $\unit[19]{mA}$  yielding a homogeneous magnetic field $\left |\vec{B}_0\right|$ of $\approx\unit[1]{\mu T}$ in the center of the coil.
The $^3$He optical pumping light from the $\unit[1083]{nm}$ laser was brought into the chamber by an optical fiber, after which the beam was expanded by a telescope in order to illuminate the whole accessible cross section of the $^3$He cell.
The light was circularly polarized with a polarizing beamsplitter cube followed by a $\lambda/4$ plate.
After traversing the $^3$He cell the pump beam was back-reflected by a mirror for a second passage through the cell, thus increasing the pumping efficiency.
The fluorescence from a $^3$He reference cell located outside BMSR-2 was used to monitor and manually adjust the laser wavelength during optical pumping.

The 894~nm light for operating the 8 CsOPMs was delivered to the sensors by eight multimode fibers.
A board containing  8 transimpedance amplifiers (mounted inside BMSR-2) pre-amplified the photodiode currents, and the ensuing voltage signals were transmitted to the data acquisition (DAQ) system located outside of the chamber.
The raw signals as well as the demodulated PD signals from a set of six CsOPMs were recorded simultaneously by (nominally) identical DAQ channels.
The remaining two CsOPM signals were recorded by a separate DAQ system and were not used for the analysis presented here.
The timebase of the DAQ system was referenced to a rubidium atomic clock  (SRS PRS10).

\begin{figure}
\hspace*{-5mm}
\includegraphics[width=1.05\columnwidth]{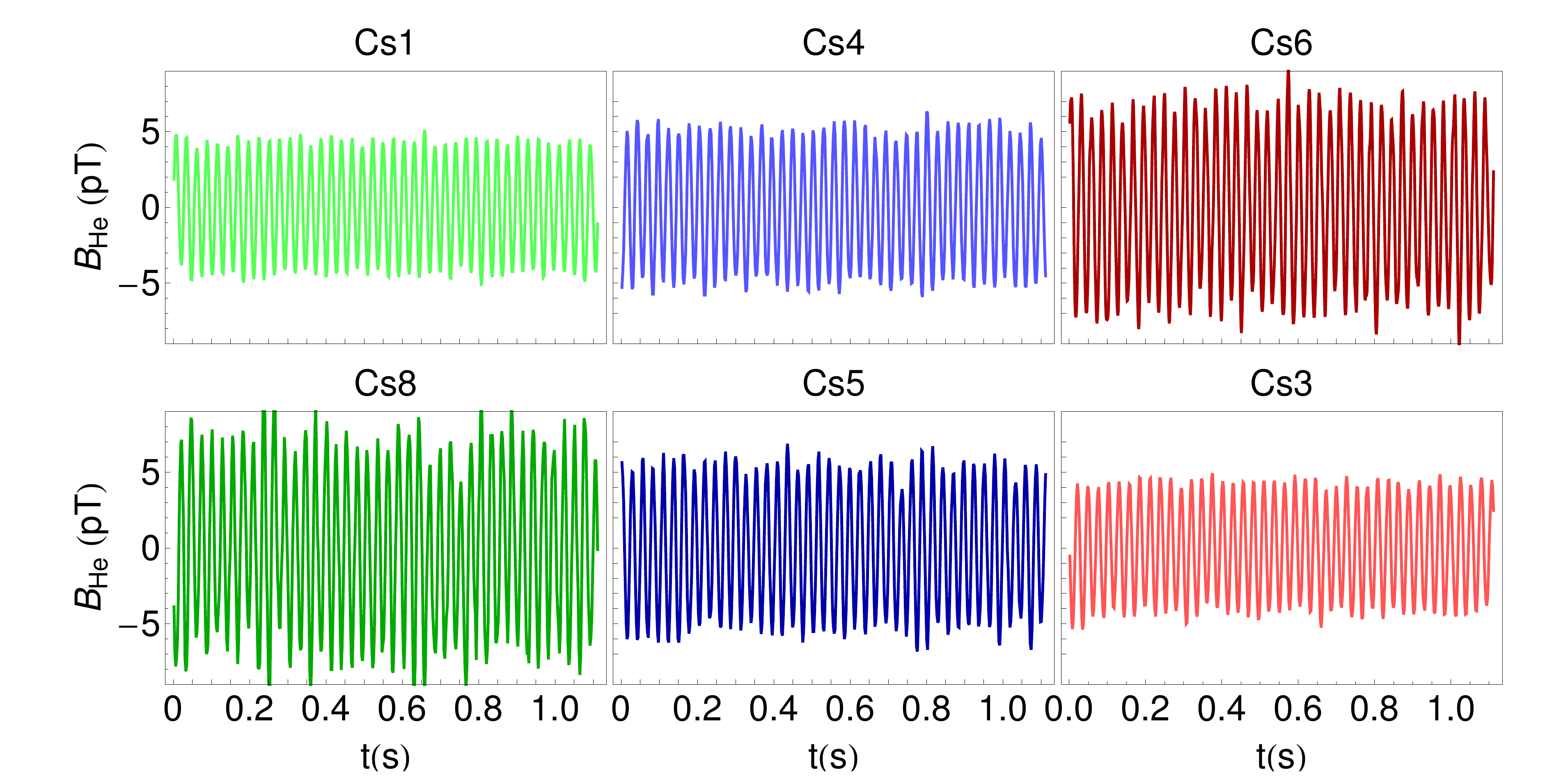}
\caption[Short Caption]{Raw phase signals of the six CsOPMs used in the analysis, rescaled to magnetic units. All subplots have identical amplitude and time scales. The $\sim\unit[36]{Hz}$ oscillation originating from the $^3$He FSP is clearly visible. The amplitudes and signal to noise ratios differ due to individual bandwidth limitations of the individual CsOPMs. Note that the amplitudes do not reflect the true magnitude of $B_\mathrm{He}$.}
\label{fig:meas_phase}
\end{figure}

Figure \ref{fig:meas_phase} shows the demodulated phase signals of six CsOPMs after rescaling to magnetic field units.
The $\sim\unit[36]{Hz}$ oscillation from the $^3$He FSP is clearly visible with no additional filtering applied to the phase data.
If not specified otherwise, the $\unit[36]{Hz}$ precession signals from the $^3$He atoms will be referred to as FSP signals in what follows.
The FSP signals in Fig.~\ref{fig:meas_phase} are clearly of varying quality.
While Cs1 (upper left in Fig.~\ref{fig:meas_phase}) exhibits a relatively good signal to noise ratio (SNR),  Cs8 (lower left in Fig.~\ref{fig:meas_phase}) is obviously performing much worse.
These differences are due to different Cs-cell qualities and CsOPM drive parameters.
A criterion to quantify the sensors intrinsic sensitivity limitation taking cell quality and drive parameters into account is the noise equivalent magnetic field (NEM) (c.f.~Sec.~\ref{sec:sn_sens} for details).
While for Cs1 we find a quite good value of $\mathrm{NEM}_{Cs1}\approx\unit[34]{fT/\sqrt{Hz}}$, Cs8 yields $\mathrm{NEM}_{Cs8}\approx\unit[158]{fT/\sqrt{Hz}}$.
For the measurements presented here, Cs1 was always performing significantly better than the other CsOPMs.
While the drive parameters can be optimized during a measurement, preselection of high quality Cs cells for the combined magnetometer is of crucial importance.
% 
%%%%%%%%%%%%%%%%%%%%%%%%%%%%%%%%%
\section{Data analysis}
\label{sec:anal}
%%%%%%%%%%%%%%%%%%%%%%%%%%%%%%%%
\noindent The data were analyzed off-line using dedicated Mathematica \cite{mathematica} codes.
\subsection{Relaxation time}
\label{sec:relax}
\begin{figure}
\hspace*{-12mm}
\includegraphics[width=1.25\columnwidth]{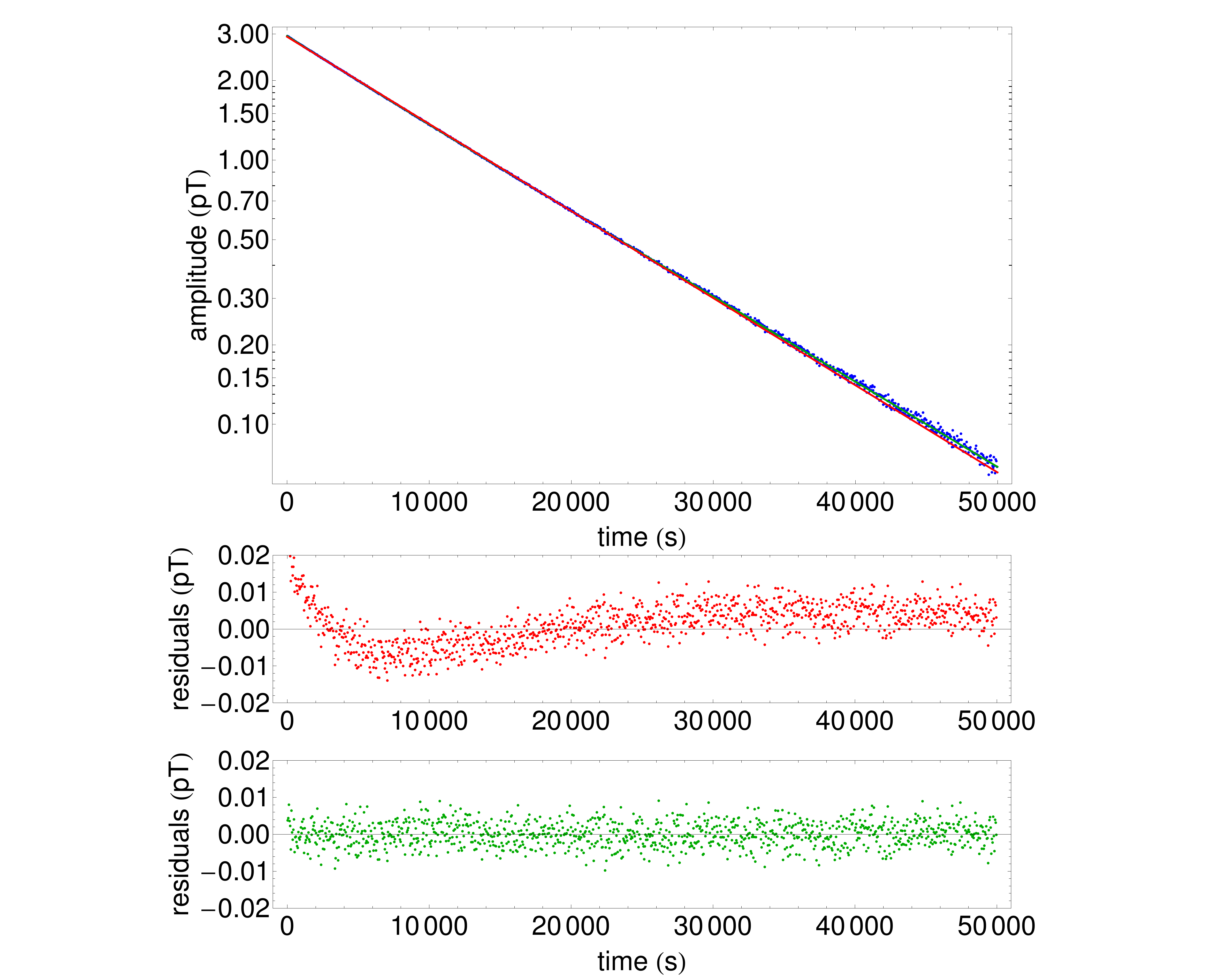}
\caption{(Color online) Measurement of $^3$He $T_2$ time. Fit of single exponentially decaying function Eq.~\eqref{eq:anal_decay2} and combined function Eq.~\eqref{eq:anal_decay3}, taking into account the relaxation of magnetic field gradients (top). The fit residuals are shown for the simple function (middle, red) and the more complex one (bottom, green). Note that the amplitudes -as in Fig.~\ref{fig:meas_phase}- do not reflect the true magnitude of $B_\mathrm{He}$.}
\label{fig:anal_t2}
\end{figure}

\begin{figure}
\hspace*{-5mm}
\includegraphics[width=\columnwidth]{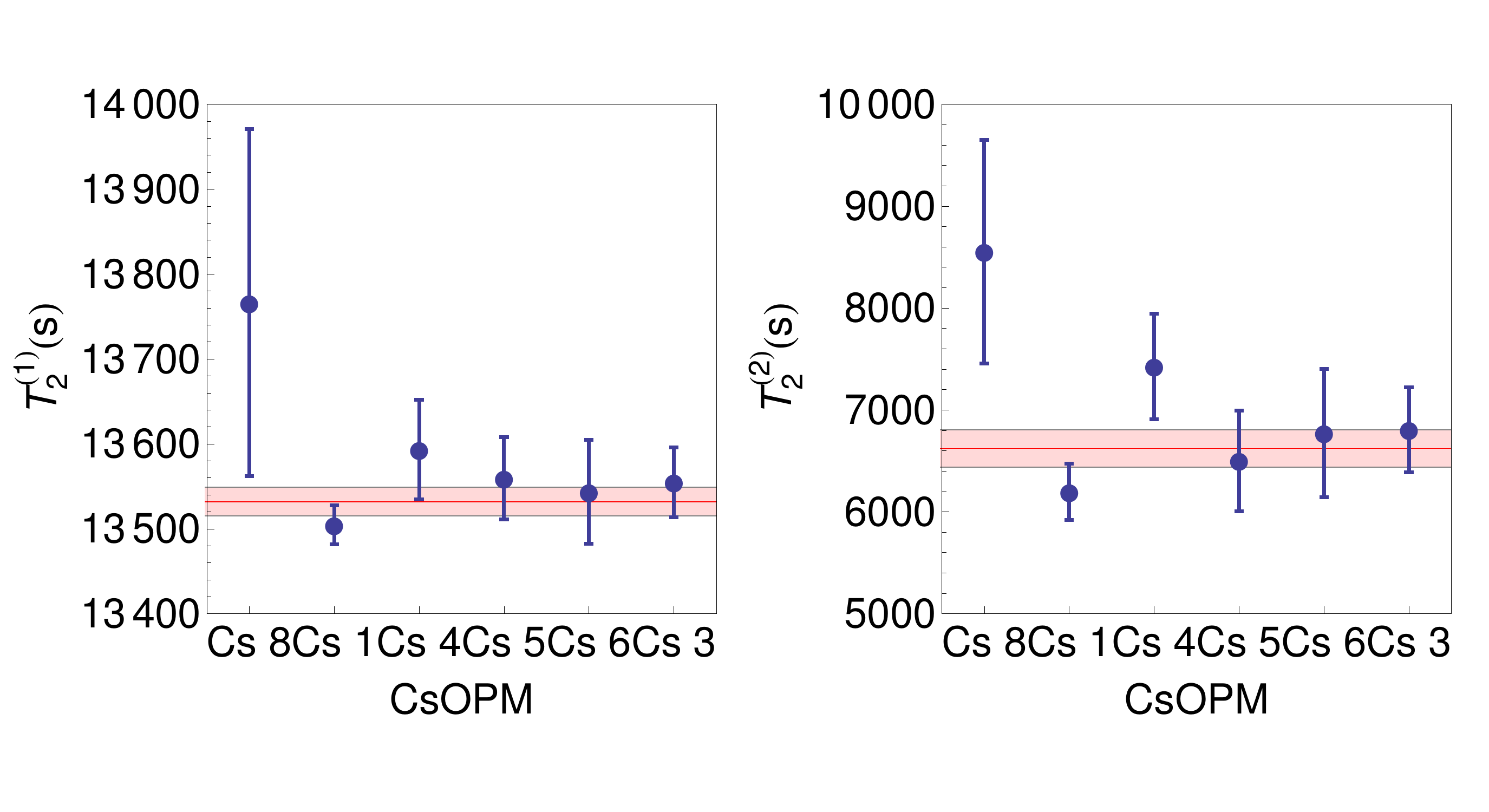}
\caption{(Color online) Decay times of $^3$He spin polarization simultaneously measured by multiple CsOPMs. The red horizontal lines denote the weighted mean and the shaded band around it gives the 1$\sigma$ confidence region. Compare Eq.~\eqref{eq:anal_decay3} for meaning of variables.}
\label{fig:decaytimes}
\end{figure}

%%%%%%%%%%%%%%%%%%%%%%%%%%%%
%
\noindent As discussed in \cite{Cates:RelInhom}, the transverse spin relaxation time $T_2$ of nuclear spin polarized $^3$He atoms is strongly affected by the presence of magnetic field gradients.
The very long $T_2$ time that can be achieved in high quality glass cells is, in general, limited by field inhomogeneities.
We recorded $^3$He FSP signals for time periods of slightly more than 10 hours.
The data, recorded at a sampling rate of $\unit[450]{Hz}$, was split into $\unit[44]{s}$ long segments over which the FSP amplitude can be assumed to be constant.
Each subset was then individually fitted using a sinusoidal function
\begin{equation}
f(t)=a_{\mathrm{off}}+a(t)\,\sin(\omega_{He} t+ \phi)\,,
\label{eq:anal_sinusoid}
\end{equation}
where $a_\text{off}$ is the offset field at the individual sensor's position and $a(t)$ the constant FSP amplitude for the respective subset.
The time dependence of these amplitudes for a single CsOPM is shown in a semi-logarithmic plot in Fig.~\ref{fig:anal_t2} that illustrates the exponential character of the decay.
The FSP signal is described by
\begin{equation}
\textcolor{black}{s(t)=a_0\,e^{-t/T_2}\,\sin(\omega_{He} t+\phi)\,.}
\label{eq:anal_decay}
\end{equation}
From a fit of the data shown in Fig.~\ref{fig:anal_t2} (from a single CsOPM) by the function 

\begin{equation}
\textcolor{black}{a(t)=a_0\,e^{-t/T_2}}
\label{eq:anal_decay2}
\end{equation}
we infer a decay time $T_2$  of $\unit[13173(4)]{s}\approx\unit[3.6]{h}$.
Closer inspection of the fit residuals reveals a small imperfection of the fit (red curves, middle graph of Fig.~\ref{fig:anal_t2}). 
This can be explained by variations of magnetic field gradients induced by the magnetic relaxation of the $\mu$-metal shield (essentially the innermost layer) after closure at the beginning of the measurement. 
This effect has already been observed in BMSR-2 and was described in \cite{Tullney13}.
An alternative fit function,

\begin{equation}
\textcolor{black}{a(t)=a_1\,e^{-t/T_2^{(1)}}+a_2\,e^{-t/ T_2^{(2)}}}
\label{eq:anal_decay3}
\end{equation}
empirically takes into account the shield relaxation by introducing a second time constant $ T_2^{(2)}$.
The fit of Eq.~\eqref{eq:anal_decay3} yields Gaussian-distributed residuals and a decay time $T_2^{(2)}$  of $\unit[13505(23)]{s}$ (green curve, lower in Fig.~\ref{fig:anal_t2}).
The same analysis, performed with all six simultaneously running CsOPMs yields the decay rates visualized in in Fig.~\ref{fig:decaytimes}.
As expected, the values for $T_2^{(1)}$ (and $T_2^{(2)}$ respectively) measured by different CsOPMs agree within their uncertainties.
The mean decay times of the $^3$He polarization, calculated from all six CsOPM measurements, are $T_2^{(1)}=\unit[13532(17)]{s}$ and $T_2^{(2)}=\unit[6621(183)]{s}$ respectively.
Although much longer decay times have been reported in the literature (see, e.g., \cite{heil2013spin}), the value achieved here is largely sufficient in the context of the present study.

\begin{figure}
\hspace*{-13mm}
\includegraphics[width=1.25\columnwidth]{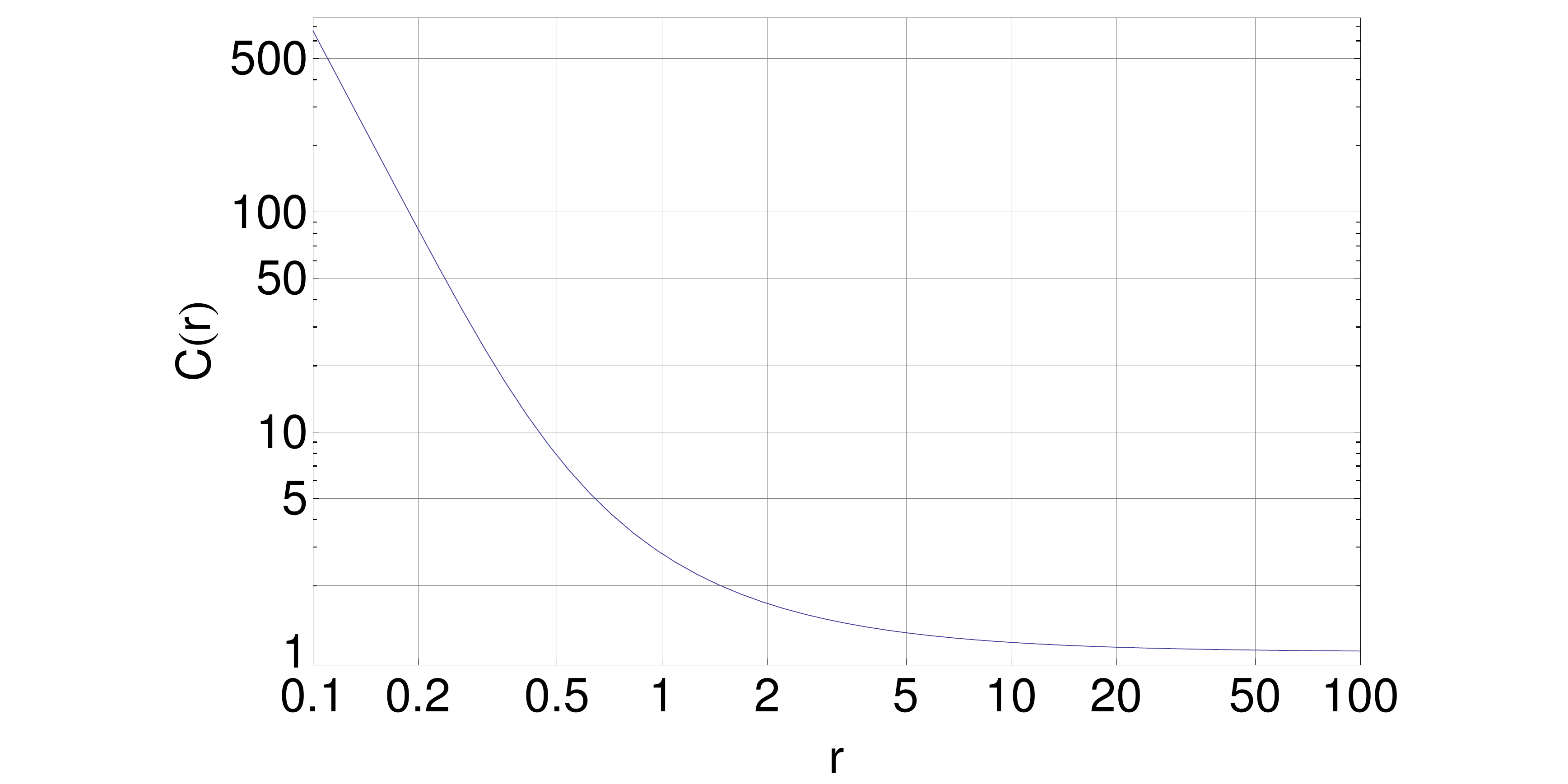}
\caption{Plot of CRLB degradation function $C(r)$ from Eq.~\eqref{eq:enhancefactor_2} for a damped sine wave as a function of $r=T_2/T_M$, the ratio of decay to measurement time.}
\label{fig:enhance}
\end{figure}

In a constant magnetic field the $^3$He FSP can be represented by a decaying single tone oscillation as introduced in Eq.~\eqref{eq:anal_decay}.
Assuming that the data have only white Gaussian noise $\mathcal{G}$ and neglecting the shield relaxation described above, we can model the experimental signal as a discrete time series of equi-spaced data points
\begin{equation}
\textcolor{black}{S_n=a\,e^{n\,T/T_2}\,\sin(\omega\,n\,T+\phi_0)+\mathcal{G}(n)\,,}
\label{eq:discrete}
\end{equation}
where $T=(\mathrm{f}_{sr})^{-1}$ is the inverse of the sampling rate $\mathrm{f}_{sr}$, i.e., the spacing between consecutive points in the time series and $\mathcal{G}(n)$ the Gaussian noise contribution to the n-th data point.
The noise is completely characterized by its power spectral density $\rho^2$ or variance $\sigma^2_{\mathcal{G}}=\rho^2 f_{bw}$ where $f_{bw}=f_{sr}/2$ is the bandwidth of the measurement.
The precision of the frequency determination of such a coherent signal over a given measurement time $T_M$ is fundamentally limited as described by information theory.
It has first been studied by Cram\'er and Rao \cite{Rao:1945:CRLB}, \cite{Cramer:1946:CRLB} who derived a lower bound for the frequency estimation variance of a signal of constant amplitude (CRLB)\cite{Rife:1974:STP}.
The corresponding bound for a damped oscillation was derived in \cite{Gemmel2010} and reads
\begin{equation}
\textcolor{black}{\sigma^2_\mathrm{f}\geq\frac{6}{(2\pi)^2\,\mathrm{SNR}^2\,T^3_M}\cdot C(T_M,T,T_2)\,,}
\label{eq:crlb_damp}
\end{equation}
with
\begin{align}
&\textcolor{black}{C(T_M,T,T_2)=}\notag\\
&\textcolor{black}{\frac{T^3_M}{6\,T^3}\cdot\frac{(1-e^{-2T/T_2})^3\,(1-\alpha)}{e^{-2T/T_2}(1-\alpha)^2-(T_M/T)^2\,
\alpha(1-e^{-2T/T_2})^2}}
\label{eq:enhancefactor}
\end{align}
where
\begin{equation}
\textcolor{black}{\alpha=e^{-2T_M/T_2}\,.}
\end{equation}
In Eq.~\eqref{eq:crlb_damp},  $\mathrm{SNR}=a_{rms}/\rho$ represents the signal to noise-density ratio with $a_{rms}=a/\sqrt{2}$ being the rms-amplitude of the FSP signal. 
$C(T_M,T,T_2)$ is a factor that takes damping into account.
For sufficiently high sampling rates ($T\ll 2\pi/\omega_{He}$), an approximative form of Eq.~\eqref{eq:enhancefactor} that is independent of T can be found and reads
\begin{equation}
\label{eq:enhancefactor_2}
C(r)=\frac{e^{2/r}-1}{3 r^3 \cosh \left(\frac{2}{r}\right)-3 r \left(r^2+2\right)}\,,
\end{equation}
where $r=T_2/T_M$ is the ratio of the decay and measurement time.
A plot of this function is shown in Fig.~\ref{fig:enhance}.
For vanishing damping, $T_2 \gg T_M$ the factor $C(T_M,T,T_2)$ becomes unity, thus reproducing the result of \cite{Rife:1974:STP}.
Since the decay time $T_2$ in our system is $\sim\unit[13,000]{s}$, one sees from Fig.~\ref{fig:enhance} that $C$ is $\approx$1 for measurement times up to several thousand seconds and can thus be neglected.
It turns out that up to integration times of several hundreds of seconds the fit results do not differ when the decay is neglected.
This justifies the use of the simple fit function \eqref{eq:anal_sinusoid} up to relatively long integration times for which the signal amplitude can be considered constant.

%%%%%%%%%%%%%%%%%%%%%%%%%%%%%%%%%%%%%%%%%
\subsection{Common noise suppression}
\label{sec:common}
\begin{figure*}
\hspace{-15mm}
\includegraphics[width=2.2\columnwidth]{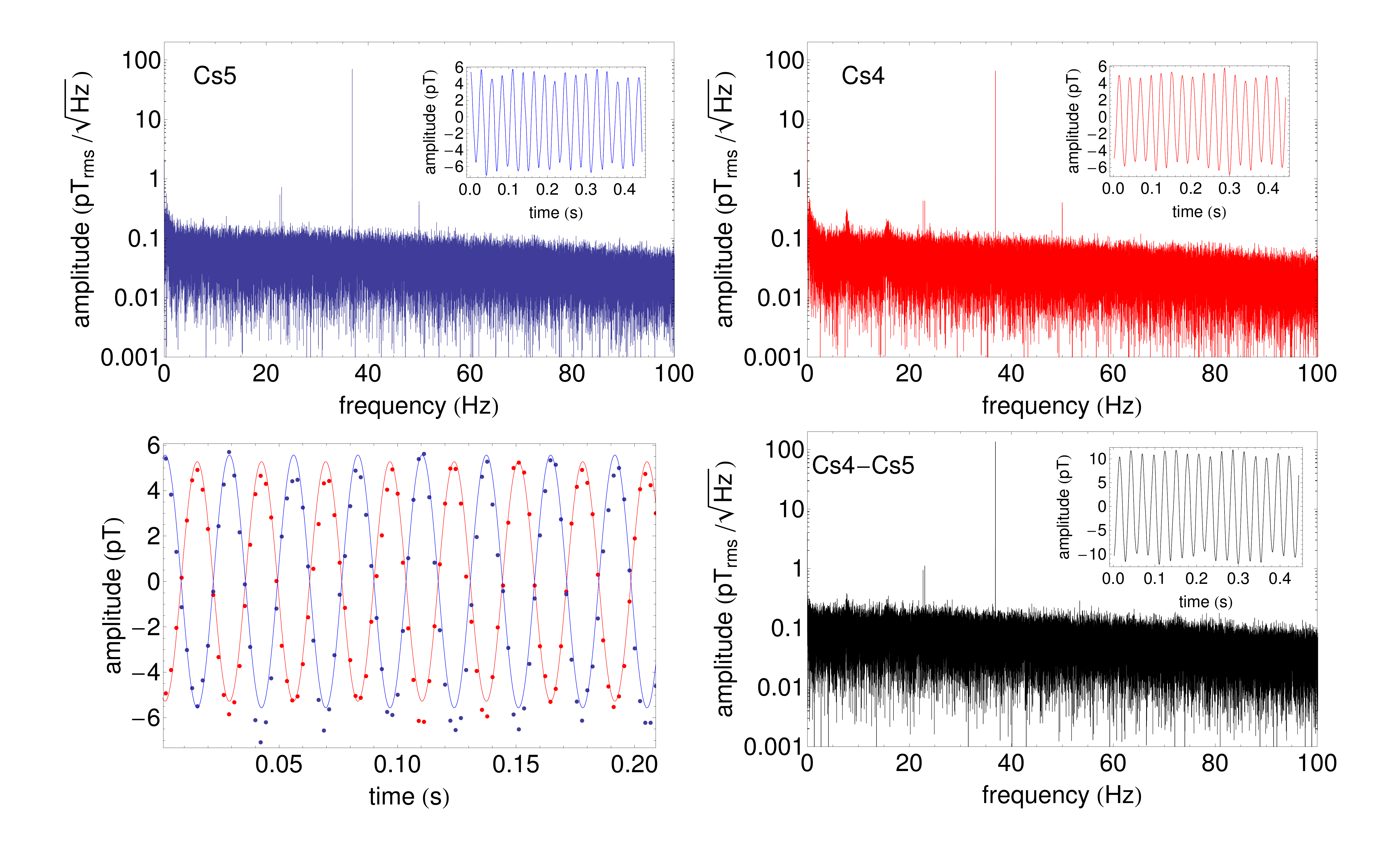}
\caption{Upper row: Fourier spectra of two CsOPMs dephased by $\pi$. Perturbations from the $\unit[50]{Hz}$ line frequency and around $\sim\unit[22]{Hz}$ are visible. Lower row: Time series of both signals, dephasing is visible (lower left), Fourier spectrum of differential signal. The $\unit[50]{Hz}$ pertubation has vanished but the $\sim\unit[22]{Hz}$ perturbations persist and are even increased (lower right). Each Fourier spectrum contains $\unit[700]{s}$ of measurement data.}
\label{fig:gradio}
\end{figure*}

%%%%%%%%%%%%%%%%%%%%%%%%%%%%%%%%%%%%%%%%%
\noindent Noise suppression effects in differential CsOPM signals as described in Sec.~\ref{sec:prepareDetect} were investigated.
For this, the phase signal from one CsOPMs (Cs5) was subtracted from a second CsOPM signal that is, by construction, dephased by $\pi$ (Cs4).
Both signals carry perturbations from the $\unit[50]{Hz}$ line frequency, as shown in the Fourier spectra in the top row of Fig.~\ref{fig:gradio}.
These perturbations are in-phase on both signals, whereas the $^3$He FSP signal is dephased by $\pi$, as visible in the lower left plot of Fig.~\ref{fig:gradio}.
In the differential signal, the $\unit[50]{Hz}$ perturbation has vanished, as evidenced by the Fourier spectrum on the lower right of Fig.~\ref{fig:gradio}.
It is expected that the (random) noise amplitude spectral densities of the two signals $\rho_4=\unit[48]{fT/\sqrt{Hz}}$ and $\rho_5=\unit[59]{fT/\sqrt{Hz}}$ add quadratically in the differential signal  $\rho_{\mathrm{diff}}=\sqrt{\rho_4^2+\rho_5^2}$, assuming no correlation between the white noise contribution to both signals.
One also expects the signal amplitudes $a_4=\unit[3.58]{pT_{rms}}$ and $a_5=\unit[3.86]{pT_{rms}}$ to add in the combined signal, $a_\mathrm{diff}=a_4+a_5$.
The expected signal to noise ratio of the differential signal can thus be written as
\begin{equation}
\label{eq:gradio}
\mathrm{SNR}_\mathrm{diff}=\frac{a_4+a_5}{\sqrt{\rho_4^2+\rho_5^2}}\,,
\end{equation}
yielding $\mathrm{SNR}_\mathrm{diff}\approx\unit[98]{\sqrt{Hz}}$ for the values given above.
Analysis of a Fourier spectrum of the differential signal shows that only a reduced SNR of $\approx\unit[85]{\sqrt{Hz}}$ is observed in the signal.
A closer analysis reveals that the amplitudes add, as expected, to $\approx\unit[7.43]{pT}$, while the noise of the differential signal is increased more than expected to a level of $\approx\unit[87]{fT/\sqrt{Hz}}$.
This discrepancy can be explained by non-Gaussian perturbations which are correlated between the two signals.
The presence of such noise components is witnessed by the double-peak around $\sim\unit[22]{Hz}$ in the Fourier spectra of Fig.~\ref{fig:gradio}.
Such perturbations may originate from higher order magnetic field fluctuations (gradient oscillations) that affect the individual sensors located in different spatial positions differently, but in a correlated manner.
Such noise processes are not suppressed by the gradiometer, their amplitudes might even add up in the combined signal, as visible in Fig.~\ref{fig:gradio} (lower right).
Nevertheless, the suppression of the common noise component was successfully demonstrated.
This technique becomes an important tool when in-phase perturbations are strong and may lead to systematic errors in the extraction of the $^3$He Larmor frequency.

%%%%%%%%%%%%%%%%%%%%%%%%%%%%%%%%%%%%%%%%%%
\subsection{Magnetic field measurements}
\label{sec:field_meas}
%%%%%%%%%%%%%%%%%%%%%%%%%%%%%%%%%%%%%%%%%%
%
\noindent In order to demonstrate the performance of the combined magnetometer we measured the magnetic field in the BMSR-2 chamber by analyzing 
consecutive 100 second long time series of the continuously recorded phase signal.
We analyzed data from the six CsOPMs that have recorded simultaneously the same $^3$He FSP.
The average frequency of each time series and its standard error were extracted by fitting Eq.~\eqref{eq:anal_sinusoid} to the data.
Special care was taken to ensure that the fit routine correctly estimates the standard errors.
Since the phase data undergoes filtering in the lock-in amplifier, the noise might not be purely Gaussian anymore.
To prevent the fit routine from underestimating the error, the variance was extracted from a Fourier spectrum of each dataset and explicitly imposed on the fit. 
The magnetic field $B_i$ and its uncertainty $\Delta B_i$ were calculated for each CsOPM using \eqref{eq:principle_larmor}.
The weighted average $\widetilde B$ and its uncertainty $\Delta \widetilde B$ were calculated for the readings of all individual CsOPMs according to
\begin{equation}
\widetilde B=\frac{\sum{\frac{B_i}{\Delta B_i^2}}}{\sum{\frac{1}{\Delta B_i^2}}}
\label{eq:avg}
\end{equation}
with standard error
\begin{equation}
\Delta \widetilde B=\left(\sum{\frac{1}{\Delta B_i^2}}\right)^{-1/2}\,.
\label{eq:avg_err}
\end{equation}
Equations~\eqref{eq:avg} and \eqref{eq:avg_err} are only valid assuming a constant magnetic field which is justified for $\unit[100]{s}$ subsets of data as shown in Sec.~\ref{sec:compare}.
The uncertainities $\Delta B_i$ of the field estimations from  data of an individual CsOPM range between 100 and $\unit[350]{fT}$ in 100 seconds, depending on the signal to noise ratio of the respective CsOPM and the field fluctuations during the measurements.
The error $\Delta\widetilde B$ of the weighted mean is typically well below $\unit[70]{fT}$ for these 100 second time slices.

\begin{figure}
\hspace*{-12mm}
\includegraphics[width=1.24\columnwidth]{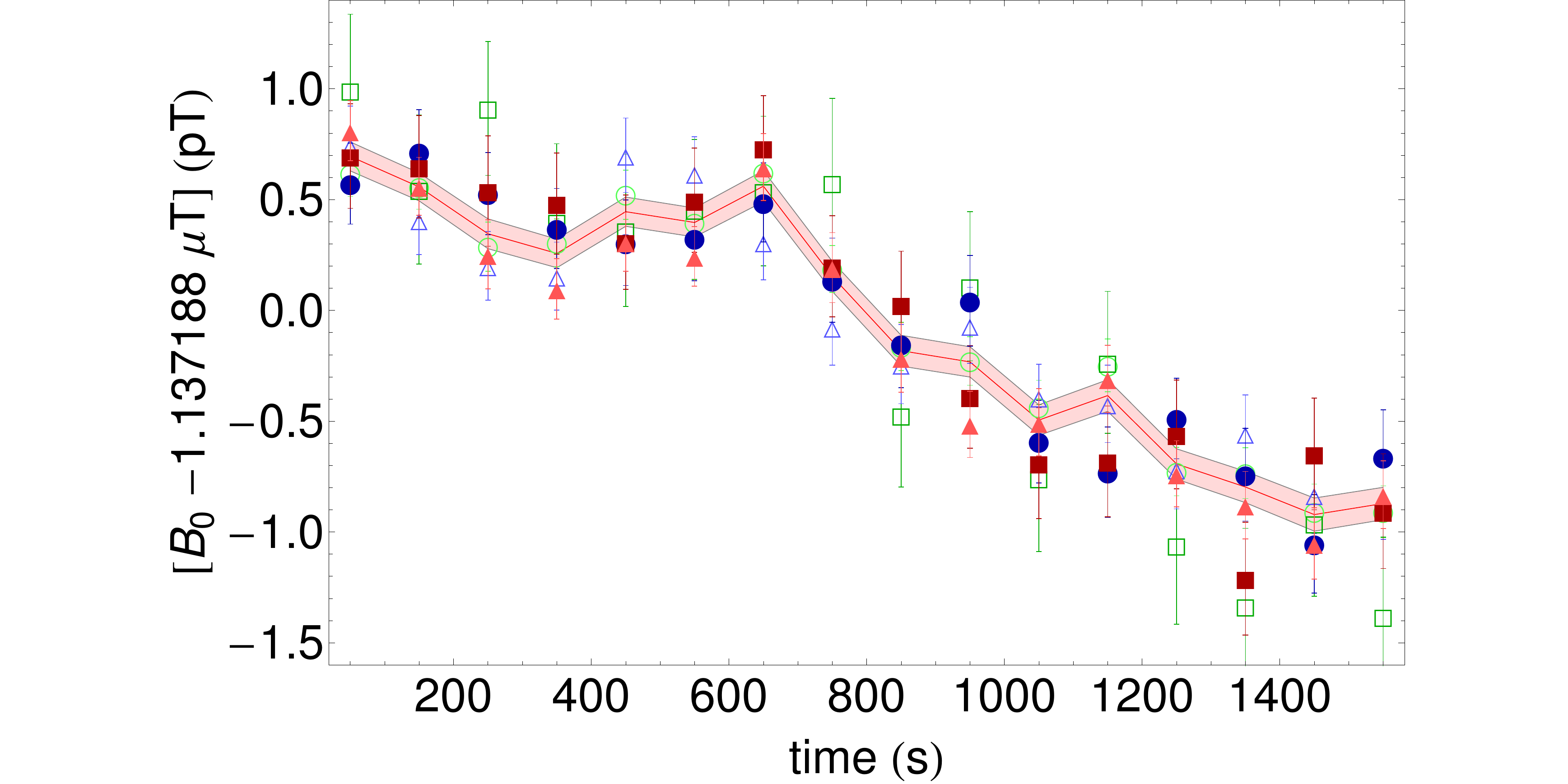}
\caption[ShortCaption]{(Color online) Time evolution of the magnetic field in the BMSR-2 chamber at PTB, estimated from consecutive 100~s subsets from data segments of six CsOPMs simultaneously detecting the same $^3$He FSP. Each plot point displays an individual CsOPM reading $B_i$ and its uncertainity $\Delta B_i$. The common ordinate is shifted by $\sim\unit[1.1]{\mu T}$. The solid line (red) and the shaded band represent $\widetilde B$ and its corresponding $1\sigma$ confidence values $\widetilde B\pm\Delta\widetilde B$.}
\label{fig:meas_time}
\end{figure}
Figure~\ref{fig:meas_time} shows, that the magnetic field in the \mbox{BMSR-2} chamber drifts by several pT on a time scale of $\approx\unit[1000]{s}$.
We will see below that these instabilities limited the determination of the combined magnetometer's sensitivity.

\subsection{Experimental determination of sensitivity}
\label{sec:compare}

The noise of any magnetometer signal will contain in general contributions from both technical and intrinsic (fundamental) noise sources.
In the best possible case the technical noise is due to the instability of the applied magnetic field.
The two sources of noise cannot be distinguished in general.
In this study we have operated simultaneously four different types of magnetometers, viz.,  the $^3$He read out by CsOPMs, the $^3$He read out by SQUIDs, the CsOPMs alone, and the SQUID(s) alone.
We can make use of this fact to distinguish noise processes that are common to all sensors from processes that affect the individual sensors.
When different types of magnetometers are operated in the same magnetic field, the technical noise contributions resulting from field fluctuations should be correlated, except for instabilities of the magnetic field gradients, since the different magnetometers are located at different spatial positions.
The Allan standard deviation (ASD) \cite{sullivan1990characterization} is a powerful tool to examine noise processes and signal stability.
A detailed account of its application to magnetometric measurements is given by Groeger et al.~\cite{Groeger:2006:SCB}.

\begin{figure}
\hspace*{-6mm}
\includegraphics[width=1.0\columnwidth]{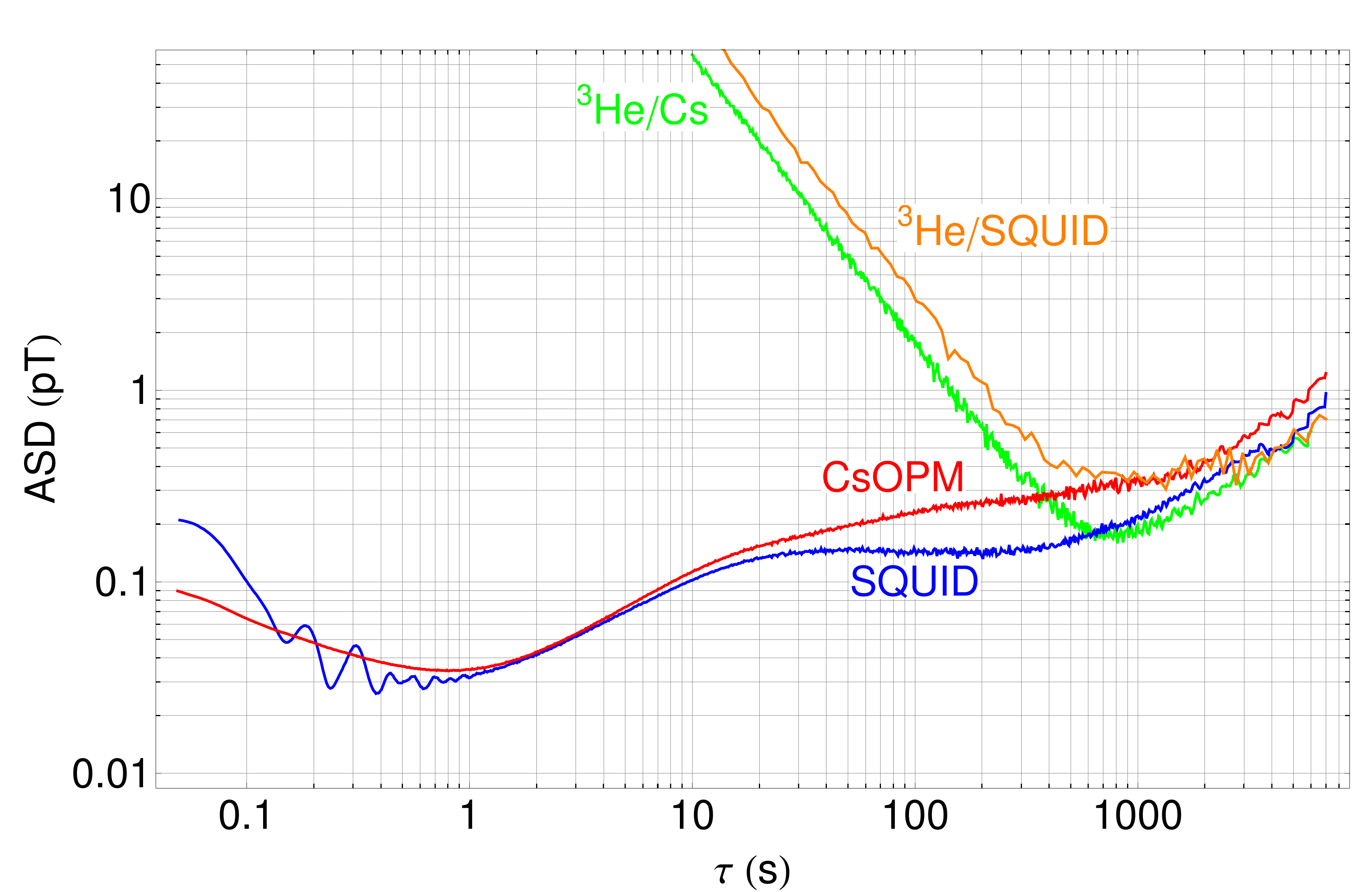}
\caption{(Color online) Allan standard deviation of simultaneous field measurements with the $^3$He/Cs (green), the  CsOPM (red), the SQUID (blue) magnetometer, and the $^3$He/SQUID (orange) respectively. For long integration times the sensitivity of all magnetometers is limited by a common noise process. The oscillatory component of the SQUID data is most probably due to mechanical vibrations of the setup. Note that the $^3$He/Cs curve lies deeper than the $^3$He/SQUID only due to the lower SNR of $^3$He/SQUID which is caused by the larger distance to the $^3$He cell.}
\label{fig:anal_asd}
\end{figure}

We have extracted the magnetic field from the $^3$He / SQUID data using the same procedure as for the $^3$He/Cs signal analysis, i.e.,  by fitting Eq.~\eqref{eq:anal_sinusoid} to the data.
The magnetometric field readings from the SQUIDs proper and from the Cs magnetometers proper were retrieved from the original data using a digital low-pass filter (bandwidth $\sim\unit[25]{Hz}$) that removes the modulation from the $^3$He FSP.
We note that these filtered magnetometer signals do not represent absolute field readings since they may be affected by detector specific offsets.
Nevertheless, as long as those potential offsets are constant, they will not affect the ASD of the signals and can thus be used for comparing the signal fluctuations.
The ASD of the measured magnetic field was calculated over a wide range of integration times for all four types of magnetometers.
Typical results are shown in Fig.~\ref{fig:anal_asd}.
The region of CRLB limited measurements up to integration times $T_M\sim\unit[500]{s}$ is restricted by the drift of the magnetic field witnessed in Fig.~\ref{fig:meas_time}.
One sees that the ASDs of all sensors, independent of the sensor type, end up on the same rising slope for sufficiently long integration times $T_M>\unit[500]{s}$.
The fact that the different ASDs do not overlap perfectly for long integration times can be explained by fluctuations of magnetic field gradients, since the different sensors were not located at exactly the same spot.
These observations support the assumption that the long-term stability of the magnetic field limits indeed the determination of the magnetometric sensitivity.
Note that for the data which entered the analysis displayed in Fig.~\ref{fig:anal_asd} the initial $^3$He polarization was $\sim\unit[20]{\%}$ smaller than for the measurements presented in Sec.~\ref{sec:field_meas}.

\begin{figure}[h]
\hspace{-10mm}
\includegraphics[width=1.2\columnwidth]{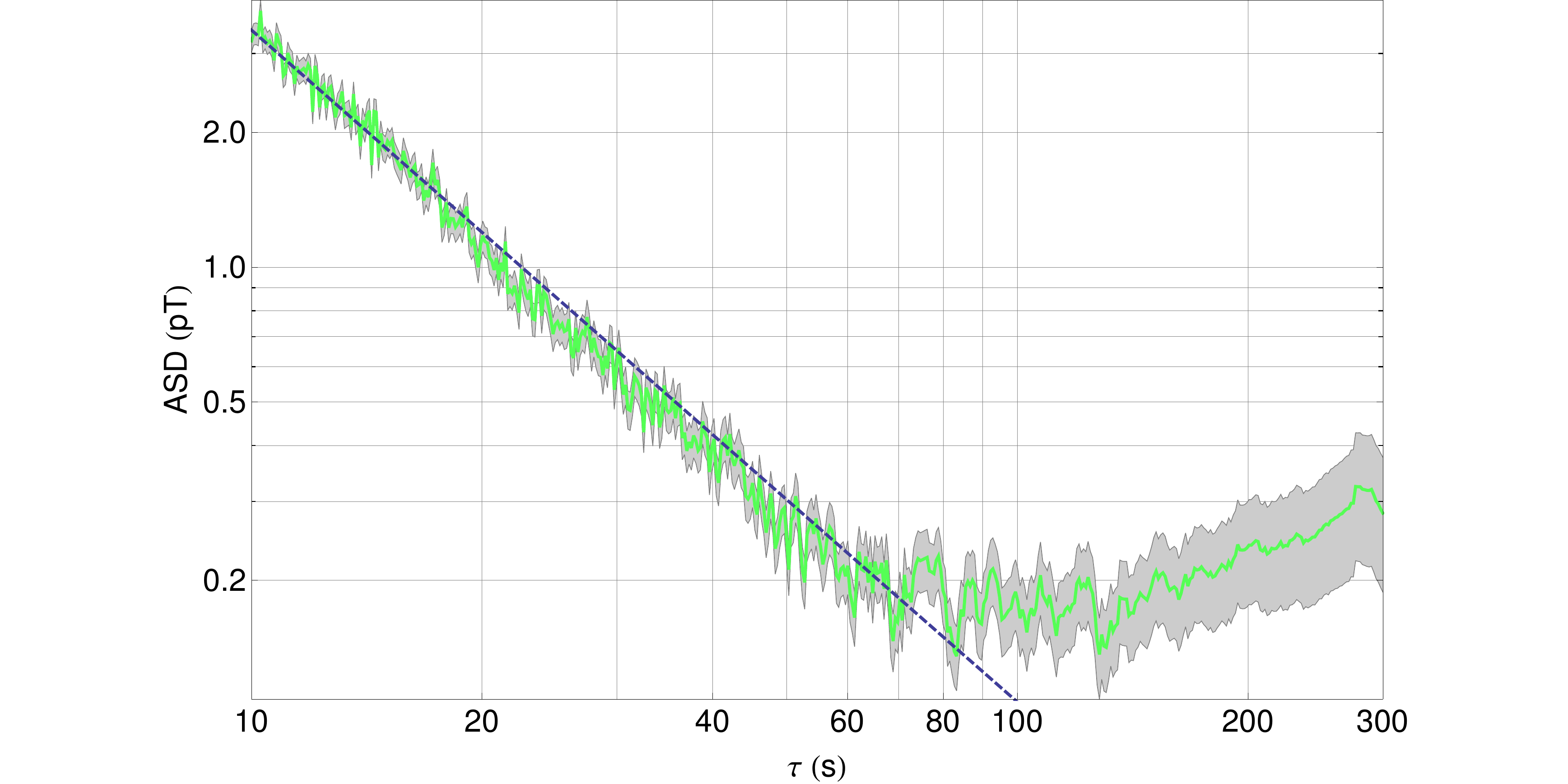}
\caption[Short Caption]{(Color online) ASD of magnetic field calculated from $\omega_\text{He}$ measured by the $^3$He/Cs magnetometer. The gray region around the curve denotes the 1$\sigma$ confidence band for each ASD point. A fit of Eq.~\eqref{eq:sens_meas} to the data up to $\tau\leq\unit[55]{s}$ is displayed as the dashed line.}
\label{fig:sensi}
\end{figure}

In order to experimentally determine the sensitivity of the $^3$He/Cs magnetometer we selected a measurement with a large $^3$He polarization and calculated the ASD of the magnetic field determined from $\omega_\text{He}$ following the same procedure as above.
The result is shown in Fig.~\ref{fig:sensi}, the shaded band giving the 1$\sigma$ confidence region for each measurement.
Where the field estimation process is CRLB limited, we find the characteristic $\tau^{-3/2}$ slope of the ASD curve, e.g., up to integration times of $\sim\unit[55]{s}$.
We determine the experimental sensitivity in this range by fitting a function
\begin{equation}
\label{eq:sens_meas}
\sigma_{B}^{\mathrm{meas}}=\eta_{B}^{\mathrm{meas}}/T_M^{3/2}\,,
\end{equation}
to the data (up to $\tau\leq\unit[55]{s}$).
The fitted function is denoted by the dashed line in Fig.~\ref{fig:sensi}.
The uncertainties of the ASD points are used as weights in this fit.
We find for the sensitivity parameter $\eta_{B}^{meas}=\unit[107.0(5)]{pT\cdot s^{3/2}}$.

We note that due to the higher $^3$He polarization compared to the data shown in Fig.~\ref{fig:anal_asd}, the ASD reaches its minimum imposed by the instability of the magnetic field already at shorter integration times.
\section{Sensitivity in shotnoise limit}
\label{sec:sn_sens}
In this section we will quantify the intrinsic noise level that is inherent to the measurement process with the given experimental parameters.
We can then compare this value to the observed noise floor to judge the quality of our measurement, allowing us to quantify the impact of technical noise (such as electromagnetic pick-up, amplifier noise,...) on the signals.
We will furthermore estimate the ultimate achievable sensitivity of the combined magnetometer under optimized experimental parameters.

The discrete nature of the charges constituting the relevant electric currents in the experiment is a fundamental source of noise.
This shotnoise is assumed to be a zero-average Gaussian process.
We calculate the intrinsic noise level of the measurement due to the shotnoise of the CsOPMs' photodiode currents and the shotnoise of the $B_0$-coilcurrent.
We refer to this fundamental noise level as the shotnoise limit in the following. 
The fundamental intrinsic noise floor of the measurement will thus have two contributions, one from the magnetic field noise, $\sigma_{field}$, and one from the CsOPM noise, $\sigma_{Cs}$. 
Since the two Gaussian sources of noise are uncorrelated one can set 
\begin{equation}
\label{eq:noise}
\sigma_{\mathcal{G}}=\sqrt{\sigma_{Cs}^2+\sigma_{\mathrm{field}}^2}\,.
\end{equation}

As described in \cite{castagna:Cells:2009}, the intrinsic noise of a CsOPM depends on the laser power, rf-power and the cell properties and can be expressed in terms of the noise-equivalent magnetic field (NEM).
For optimized rf- and laser power one obtains a minimal $\mathrm{NEM}^{\mathrm{min}}$ of $\approx\unit[12]{fT/\sqrt{Hz}}$ for the type of cells used. 
To estimate the NEM in the shotnoise limit under given experimental conditions, e.g., non-optimal rf and light power, we start from the square-root power spectral density of the photocurrent shotnoise 
\begin{equation}
\rho_{\mathrm{I},\mathrm{PSN}}=\sqrt{2e\,I_{DC}}\,,
\label{eq:pcsn}
\end{equation}
where $I_{\mathrm{DC}}$ is the measured DC-photocurrent and e the elementary charge.
The voltage shotnoise is then given by
\begin{equation}
\label{eq:VSN}
 \rho_{\mathrm{V},\mathrm{PSN}}=\rho_{\mathrm{I},\mathrm{PSN}}\,g=\sqrt{2e\,I_{DC}}\,g\,,
\end{equation}
where $g=\unit[2.53\times 10^7]{V/A}$ is the gain of the used transimpedance amplifier.
The phase change $\delta \phi$ corresponding to a voltage change $\delta U$ of the input signal is \cite{Bison:2004:DOC}
\begin{equation}
\label{eq:delpi}
\delta\phi=\frac{\delta U}{a_{Cs}}\,,
\end{equation} 
where $a_{Cs}$ denotes the amplitude of the signal after the transimpedance amplifier, and the NEM can finally be calculated using
\begin{equation}
\mathrm{NEM}=\rho_{Cs}=\frac{\delta\phi \,\Gamma_2}{\gamma_{Cs}}\,,
\label{eq:nem}
\end{equation}
with $\Gamma_2$ being the transverse relaxation rate of the Cs spin polarization.
For the best-performing CsOPM under true experimental conditions (Cs1: $\mathrm{I}_{DC}=\unit[3.25]{\mu A}$, $a_{Cs}=\unit[1.3]{V}$, $\Gamma_2/2 \pi =\unit[6.1]{Hz}$) this calculation yields 

$\mathrm{NEM}_{Cs1}\approx \unit[34]{fT/\sqrt{Hz}}$.
The photocurrents $\mathrm{I}_{DC}$ were repeatedly measured for each sensor during the experiment.
The values of $a_{Cs}$ and $\Gamma_2/2$ can be obtained by fits to the frequency sweep-responses of the individual sensors shown in Fig.~\ref{fig:response} which were also repeatedly recorded.  

The second source of noise in Eq.~\eqref{eq:noise} comes from Gaussian fluctuations of the magnetic field $B_0$ due to the coilcurrent's shotnoise. 
In the shotnoise limit $\rho_{\mathrm{field}}$ is obtained for a given coilcurrent ($I_C=\unit[19]{mA}$) and coil constant ($g_{\mathrm{Coil}}=\unit[60]{\mu T/A}$) is given by
\begin{equation}
\label{eq:csn}
\rho_{\mathrm{field}}=\sqrt{2e\,I_{c}}\,g_{\mathrm{coil}}=\unit[4.7]{fT/\sqrt{Hz}}\,.
\end{equation}
Inserting these values into Eq.~\eqref{eq:noise} leads to the shotnoise limit under the given experimental conditions of $\rho_{\mathcal{G}}^{SN}=\unit[35]{fT/\sqrt{Hz}}$.
It is obvious that  $\rho_{\mathcal{G}}^{SN}$ is dominated by $\rho_{\mathrm{Cs}}$, the contribution from the coilcurrent $\rho_{\mathrm{field}}$ playing only a negligible role.
This value can be compared to the experimentally observed noise level $\rho_{\mathcal{G}}^{exp}=\unit[27]{fT/\sqrt{Hz}}$.
In order to make this comparison consistent, we still have to consider the effect of the lock-in demodulation filter which has a transfer function $T_\text{LIA}(\omega_\text{He})=0.82$, the noise before the lock-in is thus $\rho_{\mathcal{G}}^{exp}/_\text{LIA}(\omega_\text{He})\approx\unit[35]{fT/\sqrt{Hz}}$.
We find that the observed noise floor agrees with the expected shotnoise limit.
Using Eq.~\eqref{eq:principle_larmor}, \eqref{eq:crlb_damp} and \eqref{eq:sens_meas}, we calculate the expected sensitivity parameter in the shotnoise limit from the above value of $\rho_{\mathcal{G}}^{SN}$ and the $^3$He-FSP amplitude ($a=\unit[4.15]{pT}$), yielding $\eta_B^{SN}=\sigma_{B}^{SN}(\rho_{\mathcal{G}}^{SN},a)\cdot T_M^{3/2}\approx\unit[111]{pT\cdot s^{3/2}}$.
If we compare the measured sensitivity from Eq.~\eqref{eq:sens_meas} and the shotnoise limited sensitivity we find $\eta_{B}^{meas}\approx\eta_{B}^{SN}$ .
We can thus state that the measurement was shotnoise limited.
In the same way we can compare our results to the estimated sensitivity of \cite{Cohen-Tannoudji:1969:DSM}.
We use the magnitude of the magnetic field produced by the $^3$He FSP $a_\text{CT}=\unit[6]{pT}$ reported in their paper and $\rho_{\mathcal{G}}^{exp}=\unit[27]{fT/\sqrt{Hz}}$ to find $\eta\approx\unit[77]{pT\cdot s^{3/2}}$.
This value is comparable to the estimated sensitivity of \cite{Cohen-Tannoudji:1969:DSM} given at the end of Sec.~\ref{sec:prepareDetect}, we thus conclude that the two measurements were equally sensitive.
We note that because the measurements described in \cite{Cohen-Tannoudji:1969:DSM} were done in a very weak magnetic field, the Larmor frequency is only $\omega_\text{He}/2\pi=\unit[3]{mHz}$.
At this low frequency the bandwidth limitations imposed by the readout magnetometer are not relevant.

We will now estimate the ultimate sensitivity $\eta^{min}_{B}$, under the assumption of a perfectly stable magnetic field. 
This limit is reached by maximizing $\mathrm{SNR}=a_{rms}/\rho_\mathcal{G}$ in Eq.~\eqref{eq:crlb_damp}.
We thus consider a $^3$He FSP with maximum amplitude, measured by a shotnoise limited CsOPM with minimal NEM. 
For our best paraffin-coated cells, operated under optimized conditions, a minimal NEM of $\sim\unit[7]{fT/\sqrt{Hz}}$ has been reported \cite{castagna:Cells:2009}.
The FSP amplitude is maximized for $\unit[100]{\%}$ $^3$He polarization.
We further consider a detection of the FSP without the bandwidth limitations imposed by the CsOPMs driven in the fixed frequency mode.
This could be achieved by a CsOPM driven in the phase-stabilized mode by a PLL with high bandwidth (or by a self-oscillating Cs magnetometer \cite{Budker:OM}).
For the $\sim1\mathrm{mbar}$ cell used this corresponds to $a_{rms}^{max}=41\mathrm{pT}$ at the CsOPM position.
Combining $a_{rms}^{max}$ and $\mathrm{NEM}^{min}$ from above leads to a maximum $\mathrm{SNR}$ of $\approx\unit[5800]{\sqrt{Hz}}$.
Using Eq.~\eqref{eq:crlb_damp} we thus find $\eta^{min}_{B}\approx\unit[2]{pT\cdot s^{3/2}}$ for the intrinsic sensitivity of a combined $^3$He/Cs magnetometer in this configuration.
%

%%%%%%%%%%%%%%%%%%%%%%
\section{Conclusion}
\label{sec:conc}
%%%%%%%%%%%%%%%%%%%%%%
%
\noindent We have described the design of a compact $^3$He/Cs magnetometer prototype for absolute measurements of weak magnetic fields at room temperature, and have investigated its performance.
It was shown that the magnetometer is capable of performing CRLB limited measurements within the constraints imposed by the stability of the applied magnetic field.
We demonstrated that a combined $^3$He/Cs magnetometer
consisting of a $^3$He cell and a single CsOPM as
readout-magnetometer can measure the absolute value of a $\unit[1.1]{\mu T}$
magnetic field with a standard uncertainty of $\unit[100]{fT}$ in a
measurement time of $\unit[100]{s}$, which corresponds to a relative error
below $10^{-7}$.
Measurements with simultaneous readout by multiple CsOPMs were presented and show that the standard error of the weighted mean field estimate decreases statistically with the number of CsOPMs, reaching $\Delta\widetilde B\approx\unit[60]{fT}$ in $\unit[100]{seconds}$.
This result is important because it implies that the magnetometric sensitivity of a combined $^3$He/Cs magnetometer fulfills the requirements of the current and future nEDM (and other) experiments by applying a suitable number of readout CsOPMs.
A gradiometric measurement was presented in which common-mode noise could be suppressed in the differential signal.
We estimated the intrinsic sensitivity of the prototype magnetometer for single CsOPM readout, corresponding to a standard of field estimation error of $\unit[2]{fT}$ in a
measurement time of 100~s.
For the nEDM application larger $^3$He cells are foreseen which will also increase the FSP amplitude compared to the cell used here where CsOPM- and $^3$He-cell were of similar size, leading to an even higher sensitivity.
The results show that combined $^3$He/Cs magnetometers are suitable for the absolute, high precision determination of magnetic fields in fundamental physics experiments.
\section*{Acknowledgments}
The described work was only made possible by the outstanding support from the mechanical workshops of the Physics Department at the University of Fribourg and the University of Mainz. 
This work was supported by grants from the Deutsche Forschungsgemeinschaft (HE2308/14-1) and the Swiss National Science Foundation \mbox{(200020\_140421)}.

The studies presented in this paper are part of the Ph.D. thesis of H.-C. Koch who was leading all aspects of the investigations. The other authors contributed to particular subtasks as follows:
\begin{itemize}
\item[-]\noindent A. Weis, Z. D. Gruji\'c, P. Knowles, M. Kasprzak and W. Heil were involved in the conceptual planing and construction of the
device and in the development of data acquisition methods.
\item[-]\noindent Z. D. Gruji\'c, P.Knowles, G. Bison, A. Schnabel and J. Voigt contributed to the data taking at PTB. 
\item[-]\noindent A. Weis, Z. D. Gruji\'c, G. Bison and W. Heil participated in the data analysis and the discussion of the results. 
\item[-]\noindent A. Kraft, A. Pazgalev, Z. D. Gruji\'c, P.Knowles and M. Kasprzak contributed to important precursor measurements at Johannes Gutenberg University Mainz.
\end{itemize}

% BibTeX users please use
\bibliographystyle{unsrt}
\bibliography{bib,FRAPref}
\end{document}